\documentclass{jpp}
\usepackage{graphicx}
\graphicspath{ {images/} }

\usepackage[utf8]{inputenc}
\usepackage[T1]{fontenc}
\usepackage{amsmath}
\usepackage{mathtools}
\usepackage{subcaption}
\usepackage{xcolor}

\title{Encoder-decoder neural network for solving the nonlinear Fokker-Planck-Landau collision operator in XGC}

\author{M.A. Miller\aff{1,3,4}
  \corresp{\email{mam2510@columbia.edu}},
  R.M. Churchill\aff{1},
  A. Dener\aff{2},
  C.S. Chang \aff{1},\\
  T. Munson \aff{2},
  R. Hager \aff{1}}

\affiliation{\aff{1}Plasma Physics Laboratory, 100 Stellarator Road, Princeton, NJ 08540, USA
\aff{2}Argonne National Laboratory, 9700 S Cass Ave, Lemont, IL 60439, USA
\aff{3}Columbia University, Applied Physics and Applied Mathematics Department, New York, NY 10027, USA
\aff{4}Massachusetts Institute of Technology, Department of Nuclear Science \& Engineering, Cambridge, MA 02139, USA}

\begin{document}

\maketitle

\begin{abstract}

    An encoder-decoder neural network has been used to examine the possibility for acceleration of a partial integro-differential equation, the Fokker-Planck-Landau collision operator. This is part of the governing equation in the massively parallel particle-in-cell code, XGC, which is used to study turbulence in fusion energy devices. The neural network emphasizes physics-inspired learning, where it is taught to respect physical conservation constraints of the collision operator by including them in the training loss, along with the $\ell_2$ loss. In particular, network architectures used for the computer vision task of semantic segmentation have been used for training. A penalization method is used to enforce the "soft" constraints of the system and integrate error in the conservation properties into the loss function. During training, quantities representing the density, momentum, and energy for all species of the system are calculated at each configuration vertex, mirroring the procedure in XGC. This simple training has produced a median relative loss, across configuration space, on the order of $10^{-4}$, which is low enough if the error is of random nature, but not if it is of drift nature in timesteps. The run time for the current Picard iterative solver of the operator is $O(n^2)$, where n is the number of plasma species. As the XGC1 code begins to attack problems including a larger number of species, the collision operator will become expensive computationally, making the neural network solver even more important, especially since the training only scales as $O(n)$. A wide enough range of collisionality has been considered in the training data to ensure the full domain of collision physics is captured. An advanced technique to decrease the losses further will be subject of a subsequent report.  Eventual work will include expansion of the network to include multiple plasma species.  

\end{abstract}

\section{Background}
As high performance computing (HPC) initiatives bring the advent of the exascale era, numerical simulations become capable of providing prediction with high levels of accuracy. In toroidal magnetic confinement fusion devices, the hot plasma dynamics is determined by space-time overlapping multi-scale interactions among complex multiscale physics. Calculating these interactions is computationally expensive, especially in the case of total-f gyrokinetic codes, which solve for the motion of particles on turbulence scales and the background scale together without scale separation. XGC (X-point Gyrokinetic Code) is a massively parallel hybrid Lagrangian-Eulerian particle-in-cell based gyrokinetic code focused on simulating the highly nonlinear, non-equilibrium edge region of fusion devices \citep{Ku2018}. An expensive part of the code is the collision operator acting on the five-dimensional (5D) particle distribution functions. XGC uses the nonlinear Fokker-Planck-Landau (FPL) collision operator \citep{landau1936transport}, seen in Equation \ref{eq:fpl}, on a two-dimensional normalized velocity grid, $v_\perp/v_{th},v_\parallel/v_{th}$ (spatial points are independent in the collision operator) \citep{Yoon2014,Hager2016}. At every time step, particles tracked in 3D configuration space and 2D velocity coordinates are histogrammed into velocity particle distribution functions $f$, on regular, 2D velocity grids at each configuration space vertex. An implicit Picard iteration scheme is then used to solve the integro-differential equation involving the collision operator on the 2D velocity grid at each configuration space vertex, with a typical number of configuration space vertices in a simulation on the order of $10^1$ -- $10^2$ million.
\begin{equation}
  \frac{df_a}{dt} = \sum_b C_{ab}(f_a;f_b')
    = -\sum_b \frac{e_a^2e_b^2ln\Lambda_{ab}}{8\pi\epsilon_0^2m_a} \nabla_v \cdot \int \textbf{U} \cdot \left(\frac{f_a}{m_b} \nabla_v' f_b' - \frac{f_b'}{m_a} \nabla_v f_a\right)d^3v'
    \label{eq:fpl}
\end{equation}
%
%
In Equation~\ref{eq:fpl}, $a$ and $b$ denotes separate species, although self-collisions are included in the case that $b = a$. $f_a$ and $f_b$ are the particle distribution functions of the species, $e_{a,b}$ is the charge, $m_{a,b}$ is the mass, $\ln \Lambda_{ab}$ is the Coulomb logarithm, and $\mathbf{U}$ is a tensor that is a function of the relative vector $\mathbf{u} = \mathbf{v} - \mathbf{v'}$:

\begin{equation}
    \textbf{U} = \frac{u^{2}\textbf{I} - \textbf{u}\textbf{u}}{u^{3}}
\end{equation}

The run time for the Picard iteration solver of the collision operator in XGC1 for production runs with electrons turned on is currently ~10\% of the overall compute time. This number increases in higher collisionality regimes. The run time for the operator, however, is $O(n^2)$, where $n$ is the number of plasma species. As the XGC1 code begins to attack problems including a larger number of species, such as including impurities like tungsten and its multiple charge states in devices like ITER, the collision operator will become the dominant computation \citep{Dominski2019}.

Deep neural networks have proven successful for many computer vision tasks, including image recognition, as well as text and speech recognition. Given their success, a great deal of effort has been expended in utilizing them to solve complex problems across science and engineering. Specifically in physics, their application is widespread. From learning field theories using a set of training data produced from a physical field \citep{Qin2019}, to simulating particles using a graph network by combining an ODE integrator with Hamiltonian dynamics \citep{Sanchez-Gonzalez2019}, to predicting non-linear, cosmological large scale structures of the universe \citep{He2019}, neural networks have proven incredibly versatile. 

Recognizing the importance of efficient PDE solvers across the field of physics, a more concerted effort is underway to use neural networks to help solve especially intractable PDEs. Almost always, training of a PDE solver using neural networks depends on minimizing an objective function that is composed primarily of the quadratic residual, also known as mean-squared error or $\ell_2$ loss. Consider a stationary PDE of the form:
\begin{equation}
\begin{aligned}
    &Lu = f, x\in\Omega\\
    &Bu = g, x\in\Gamma\subset\partial\Omega
    \label{eq:pde}
\end{aligned}
\end{equation}
where $L$ is a differential operator, $f$ is a forcing function, $B$ is a boundary operator, and $g$ is boundary data. Solutions then require the minimization of an objective function composed of the $\ell_2$ loss as follows:
\begin{equation}
    C = \alpha ||L\hat{u} - f||^{2}
    \label{eq:gen_l2}
\end{equation}
where $\hat{u} = \hat{u}(\theta,\phi,...)$ is the ansatz that depends on some number of trainable parameters $\theta$, $\phi$, etc. that can be optimized to minimize Equation \ref{eq:gen_l2}. Here $\alpha$ is some coefficient that can vary and that becomes especially important as other terms are added to the objective function. 

While this methodology is common to PDE neural network techniques, the incorporation of boundary conditions, or any constraints in general varies. Often, these constraints are built into the network itself as ``hard" constraints, either directly into the architecture or into a constraint on the ansatz. Some attempt to do the former by adding a differentiable PDE layer that enforces spatial constraints \citep{Jiang2020} or in the case of a simple governing PDE, the use of an untrainable ``physics embedded decoder" that replicates the differential operator and boundary conditions directly \citep{ArvindT.Mohan2017}. Others use the boundary conditions to constrain the ansatz itself and then proceed with unconstrained optimization of an objective function like in Equation \ref{eq:gen_l2} \citep{Berg2018}. One last set of techniques treats the constraints as ``soft" and instead incorporates them into the objective function as penalty terms. This is done by adding the boundary conditions directly into the loss function \citep{Sirignano2018} or by leveraging the use of a coefficient like $\alpha$ in Equation \ref{eq:gen_l2} that allows for a relative prioritization of different constraints in the loss function \citep{Beucler2019}.

This paper employs a technique very similar to the latter. Using a penalization method to optimize ``soft'' constraints, it aims to learn the nonlinear transformation of the collision operator in Equation \ref{eq:fpl}, enabling the XGC code to be accelerated.  Since the FPL collision operator can be expressed as solving the collisional change $\Delta f_a = F(f_a,f_b,...)$, where $\Delta f_a$ is the same size on the normalized 2D velocity grid as $f_a$, $f_b$, etc. in a non-equilibrium edge plasma, here we use encoder-decoder deep neural networks used in the common computer vision task of semantic segmentation.

\section{Machine Learning in Computer Vision}

Computer vision has been one of the largest drivers for machine learning algorithm development. During the ImageNet competition, it was quickly noticed that deep convolutional neural networks (CNN) performed best for image recognition \citep{Alom2017}. These networks take in an input image and output a set of scores corresponding to a set of possible categories that identify the image. The largest of these scores corresponds to the prediction of the network. Structurally, the input has dimensions H x W x 3, where H is the number of image pixels vertically, W is the same horizontally, and 3 corresponds to the three color channels, red, green, and blue. The output is simply a 1D vector with C elements, where C is the prescribed number of possible identification classes. A CNN contains a variable number of hidden layers which perform convolutions on the input image in an attempt to discern underlying features to aid in identification. At each of these layers, a convolution kernel performs a convolution, or formally a cross-correlation, on the data. For neural networks, this is performed in 2D as follows (although the operation generalizes to higher dimensions): 
\begin{equation}
    (f*g)(i,j) = \sum_{m=0}^{M} \sum_{n=0}^{N} f(m,n)g(i+m,j+n)
    \label{eq:convolution}
\end{equation}
where $f$ is the input into the convolution layer, $g$. For a function represented by a matrix, as is the case for neural networks, $m$ and $n$ simply iterate over the two dimensions of the matrix, $M$ and $N$, respectively, and $i$ and $j$ are used to index a particular element of the convolution \citep{Kao}.

Along with these convolutional layers, the data is passed through a nonlinear activation function at each layer, allowing the network the flexibility to learn the nonlinearities associated with many real-world problems. Every few convolutions, the data is passed through a learned filter that ``summarizes" the features in the input image. These layers, called pooling layers, reduce the number of parameters in the network. Finally, the CNN ends with a set of fully connected (FC) layers that terminate with a linear operation that transforms the final 2D feature map into a vector containing the C class scores. As with any gradient-based optimization problem, a loss function allows for computation of gradients and updating of the weights of the network to attempt to minimize the loss \citep{Kao}.

\begin{figure}
    \centering
    \includegraphics[scale=0.35]{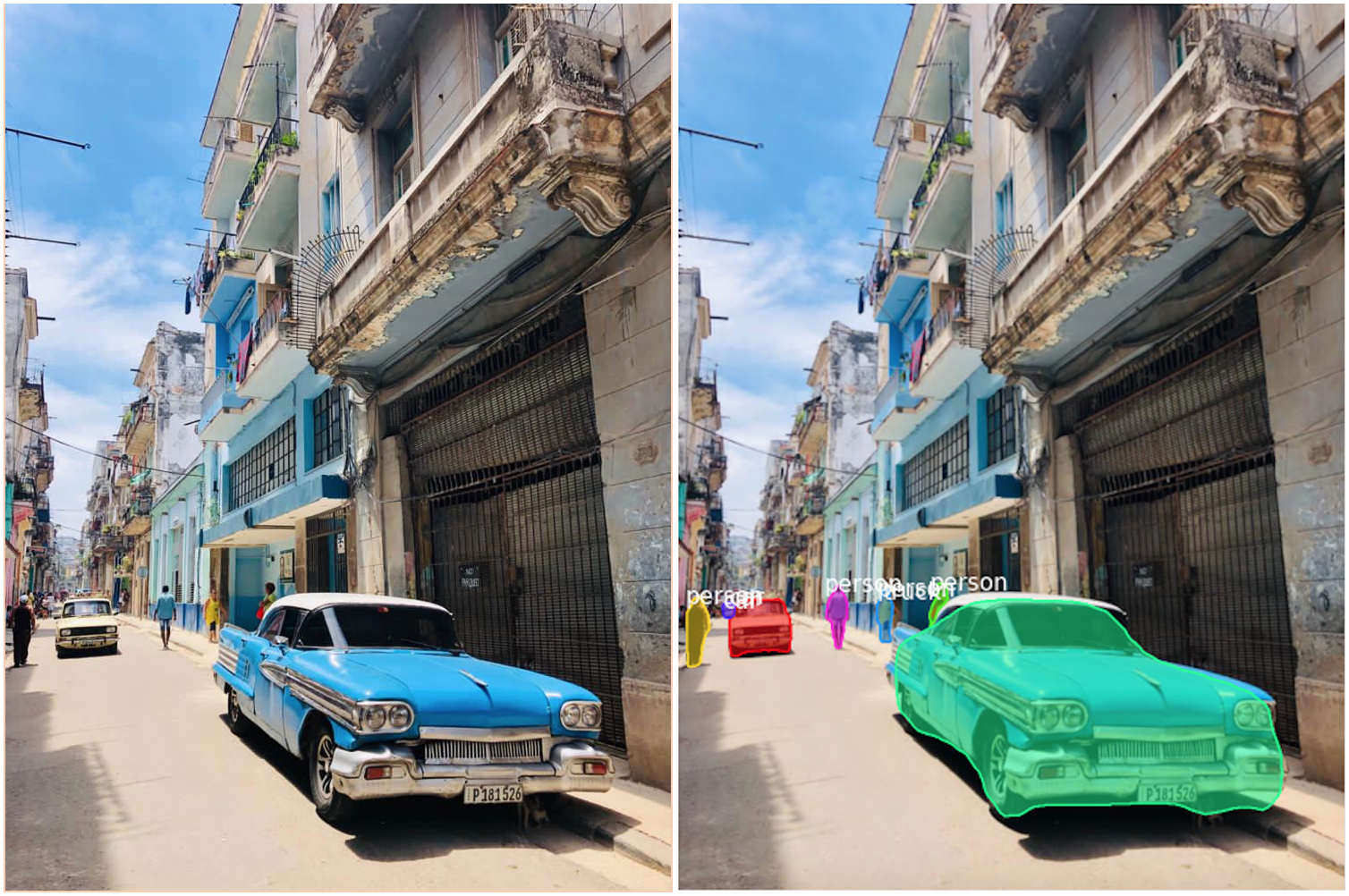}
    \caption{Example image of semantic segmentation being used to identify all cars and people in an image. Each object in the right image corresponds to a class prediction by the neural network, for the input image on the left.}
    \label{fig:semseg}
\end{figure}

As computer vision networks increased in sophistication, they turned to more complicated tasks like that of semantic segmentation. This involves classifying each pixel in an image, allowing the network to identify every part of an image. An example of an image parsed by a neural network with this goal is shown in Figure~\ref{fig:semseg}. This framework of making pixel-wise decisions on a input dataset is of great use to a variety of unrelated problems, in particular to making accelerating solutions to the equations governing the collision of particles in a tokamak. To understand why, it is necessary once again to consider the spatial arrangement of data that semantic segmentation techniques manipulate. 

Semantic segmentation is the computer vision task of labeling each pixel with the object class that pixel belongs to, e.g. marking all pixels that belong to a car in an image. For semantic segmentation, the input is an image of dimensions H x W x 3. As each pixel receives a class score, the output takes the form H x W x C. Since the image has H x W pixels, the output array must contain H x W vectors of C elements. To do this, the network "encodes" the input image by down-sampling it into a series of arrays that abstracts the features of an image, before "decoding" these features by up-sampling these feature maps back into a an array of dimensions H x W x C. The decision of what a certain object is now made at a pixel level, H x W times. For each of these pixels, there is an associated cross-entropy loss function from which the total overall loss is computed, which is then used in backpropagation to update the overall gradients.

\section{Computer Vision Technique for Collision Operator}

Semantic segmentation techniques are a good starting point for attacking the collision operator prediction problem, specifically because of the similarities between the hierarchies of convolutions in CNNs and the velocity-space gradients in the FPL collision operator. To perform the collision calculation, at every timestep, XGC performs particle to mesh interpolation, creating distribution functions on regular 2D velocity grids of size $N_{v_{perp}}$ x $N_{v_{par}}$ at each configuration grid vertex. This array can be thought of as a set of bins representing the number of particles at particular velocities, both parallel and perpendicular to the magnetic field lines.

XGC creates distribution functions on each mesh vertex for each particle type simulated. In this case, we consider electrons and a single ion species. The collision operator takes in these distribution functions and performs the collision calculation according to the FPL equations and outputs the change in the distribution function of species $a$, labeled $\Delta f_{a,col}$. For example the ion  $\Delta f_{i,col}$ is:
\begin{equation}
    \Delta f_{i,col} = \Delta t \cdot C_i(f_i,f_e) = \Delta t \cdot \left[ C_{ii}(f_i,f_i) + C_{ie}(f_i,f_e) \right],
    \label{eq:deltaf}
\end{equation}
where $\Delta t$ is the XGC ion timestep size.
This $\Delta f_{a,col}$ has the same configuration space size as the original distribution function, much like semantic segmentation techniques, which preserve the size of an input image. 

One main difference between the collision operator and typical semantic segmentation is $\Delta f_{a,col}$ is continuous, such that it's a regression problem instead of a classification problem. We use a mean-squared error loss (or $\ell_2$) as part of the loss function:
\begin{equation}
    L_{\ell_2} = \frac{1}{N} \sum_{i=1}^N (\Delta f_{XGC} - \Delta f_{ML})^2
    \label{eq:l2}
\end{equation}
where N is the number of configuration space vertices, $\Delta f_{XGC}$ is the result of the typical Picard iteration solver used by XGC, and $\Delta f_{ML}$ is the predicted $\Delta f$ from the machine learning neural network. Given that this is the only difference with semantic segmentation frameworks, networks that have proven successful in this area can be directly used to predict the outputs of the Picard iteration solver, with a minor tweak to the final layers of the network and a loss function appropriate to this problem.

\section{Physics-Inspired Constrained Learning}

In a general regression problem, the $\ell_2$  loss is sufficient to train a network to minimize the loss of a nonlinear transformation. In this problem, however, we are dealing with a physical system that is constrained by physical laws, namely the laws of mass, momentum, and energy conservation. The $\ell_2$ loss does not directly take into consideration the conservation laws of the collision operator and is often insufficient to ensure these physical constraints are satisfied. For two species, $a$ and $b$, the conservation equations can be summarized as follows \citep{Helander2002}:
\begin{equation}
    \int \phi_a(\mathbf{v}) C_{ab}(f_a,f_b) \, d^3v = -\int  \phi_b(\mathbf{v}) C_{ba}(f_b,f_a) \, d^3v
    \label{eq:cons_eq}
\end{equation}
where $\phi(\mathbf{v}) = \{m,m\mathbf{v}, \frac{1}{2}m v^2\}$ represent mass, momentum, and kinetic energy. 

For conservation of mass it can be shown more specifically that:
\begin{equation}
    \int C_{ab}(f_a,f_b) \, d^3v = 0
    \label{eq:cons_eq_mass}
\end{equation}

Note also that for self-collisions, $a=b$, Equation \ref{eq:cons_eq} reduces further:
\begin{equation}
    \int \phi_a(\mathbf{v}) C_{aa}(f_a,f_a') \, d^3v = 0
    \label{eq:cons_eq_self}
\end{equation}

\subsection{XGC Implementation}

Details of the Fokker-Planck-Landau collision operator can be found in \citep{Yoon2014} and \citep{Hager2016}. We list here mathematically the conservation properties that XGC uses in the code. For now, we will present continuous, analytical equations, and not the detailed numerical implementations (see \citep{Hager2016}). Also, we will retain the full 3D-3V notation from Section I, even though XGC is gyrokinetic (3D-2V), and there are a number of differences, but this will suffice to illustrate the main points. We drop the explicit notation inputting the distribution functions into the collision operator ($C_{ab}$, as it is redundant with the subscripts of the collision operator. For simplicity we limit ourselves to the two-species case, electrons (``e'') and a single ion specieis (``i'').

Using the following moment definitions:
\begin{equation}
\begin{array}{rcl}
    n_a & =&  \int f_a d^3v \\
    P_a & = & \int m_a \mathbf{v} f_a d^3v \\
    E_a & = & \int \frac{1}{2}m_a v^2 f_a d^3v
\end{array}
\end{equation}
The conservation properties calculated in XGC are as follows:
\begin{equation}
\begin{array}{rcl}
    \Delta n_i & = & dt \left[\int C_{ii} \, d^3v + \int C_{ie} \, d^3v \right] = 0\\
    \Delta n_e & = & dt \left[\int C_{ee} \, d^3v + \int C_{ei} \, d^3v \right] = 0\\
    \Delta P_i & = & dt \left[\int m_i \mathbf{v} C_{ii} \, d^3v + \int m_i \mathbf{v} C_{ie} \, d^3v \right]  \\
    \Delta P_e & = & dt \left[\int m_e \mathbf{v} C_{ee} \, d^3v + \int m_e \mathbf{v} C_{ei} \, d^3v \right] \\ 
    \Delta E_i & = & dt \left[\int \frac{1}{2} m_i v^2 C_{ii} \, d^3v + \int \frac{1}{2} m_i v^2 C_{ie} \, d^3v \right]  \\
    \Delta E_e & = & dt \left[\int \frac{1}{2} m_e v^2 C_{ee} \, d^3v + \int \frac{1}{2} m_e v^2 C_{ei} \, d^3v \right] \\ 
\end{array}
    \label{eq:cons_eval}
\end{equation}
These are combined to form the conservation laws (see Equation \ref{eq:cons_eq} and \ref{eq:cons_eval}):
\begin{equation}
\begin{array}{rcl}
    \Delta n_i & = & 0 \\
    \Delta n_e & = & 0 \\
    \Delta P & = & \Delta P_i + \Delta P_e = 0 \\
    \Delta E & = & \Delta E_i + \Delta E_e = 0
\end{array}
\label{eq:xgc_cons}
\end{equation}

Since these are numerical solutions, the terms in Equation \ref{eq:xgc_cons} will not be exactly 0. The quantities in Equation \ref{eq:xgc_cons} when normalized provide a test to determine how close the numerical solver is to satisfying these physical conservation properties. Equation \ref{eq:conv_crit} gives the conditions for the normalized quantities to satisfy for the algorithm to be considered converged in XGC's numerical Fokker-Planck solver. The numerical precision enforced by these criteria at each XGC ion timestep is such that the accumulated error throughout the entirety of the XGC simulation is kept below an overall simulation threshold value. The criteria, however, as they are written in Equation \ref{eq:conv_crit}, are overly stringent, and can be relaxed to up to $10^{-6}$ without sacrificing accuracy of the overall simulation.

\begin{equation}
\text{converged if} 
\begin{dcases}
    \omit\hfil$ \left| \dfrac{\Delta n_i}{n_i} \right| $\hfil & < 10^{-10}\\
    \omit\hfil$ \left| \dfrac{\Delta n_e}{n_e} \right| $\hfil & < 10^{-10} \\
  \omit\hfil$ \dfrac{\left|\Delta P_i + \Delta P_e\right|}{\left| n_i m_i v_{th,i}\right| + \left| n_e m_e v_{th,e} \right|} $\hfil & < 10^{-7} \\
    \omit\hfil$ \dfrac{\left|\Delta E_i + \Delta E_e\right|}{\left| E_i + E_e \right|} $\hfil & < 10^{-7}
\end{dcases}
    \label{eq:conv_crit}
\end{equation}
where $v_{th,i}$ and $v_{th,e}$ are local-average ion and electron thermal speeds, $v_{th} = \sqrt{T/m}$. The average electron momentum is usually negligible compared to the ion momentum.

\subsection{ML Implementation}

We enforce these physical conservation constraints in our neural network by including them as regularization terms in the loss:
\begin{equation}
    L_{cons} = \sum_{j=1}^3 \lambda_j \sum_{a = \{i,e\}} \frac{\langle  \phi_a^j \, \Delta f_{a,ML} \rangle}{A_j}
    \label{eq:cons_loss}
\end{equation}
This equation incorporates Equation \ref{eq:cons_eq} into the loss, seeking to minimize the change in these quantities due to the collision operator. Here, $\phi_a^j = \{ m_a, m_a \mathbf{v}, \frac{1}{2}m_a v^2 \}$  and $A_j = \{ n_{i}, \left| n_i m_i v_{th,i}\right| + \left| n_e m_ev_{th,e} \right| \,, \left| E_i + E_e \right| \}$ (as in Equation \ref{eq:conv_crit}) for ion density = electron density, momentum, and energy conservation, and $\langle \rangle$ represents the numerical velocity integral. The quantity $\lambda_j$ represents a hyperparameter scaling factor that allows for modification of how heavily each conservation property is weighed in the overall loss function. This in turn affects how much the network corrects for each of the conservation errors compared to the $\ell_2$ loss. Training results are rather sensitive to changes in these $\lambda_j$. For values too high, the conservation loss decreases too quickly and then plateaus. For values too low, the $\ell_2$ loss dominates optimization. As a result, tuning of these hyperparameters (in conjunction with the learning rate) allowed determination of the optimal values of $\lambda_j$. More sophisticated optimization schemes have been since been implemented. In particular, using an augmented Lagrangian method dynamically updates these hyperparameters, which lessens the burden of manually picking the $\lambda_j$ \citep{Dener2020}. The overall loss function is simply the sum of $L_{\ell_2}$ (Equation \ref{eq:l2}) and $L_{cons}$ (Equation \ref{eq:cons_loss}).

\section{Training Setup and Architecture}

\begin{figure}
    \centering
    \includegraphics[scale=0.4]{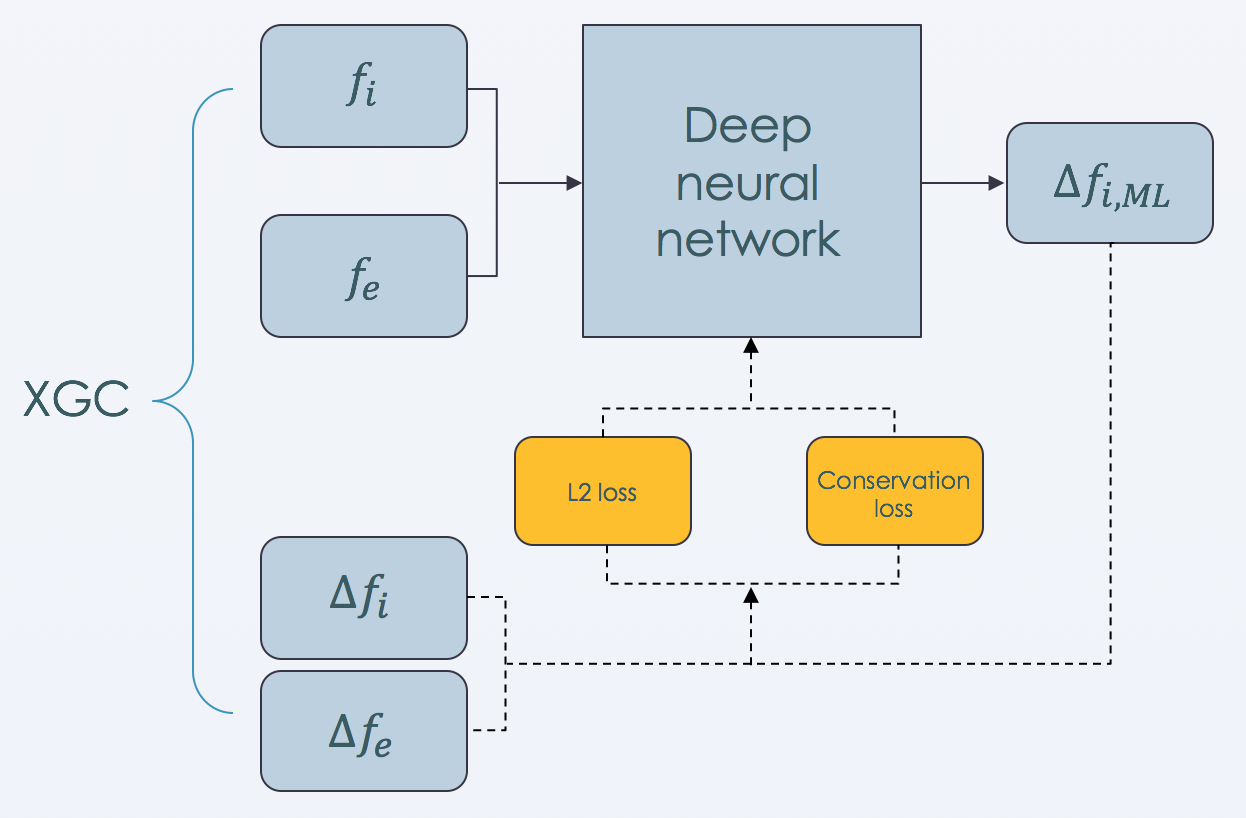}
    \caption{Schematic of the training processes used for collision prediction.}
    \label{fig:schematic}
\end{figure}

A dataset of the particle distribution functions $f_i$ and $f_e$ were gathered from XGC, and collision kernel output $\Delta f$ of the existing Picard iteration scheme was generated. These comprised over 2.5 million samples of the $N_{v_\perp}$ x $N_{v_\parallel} = 32$ x $31$ velocity-space grid distribution functions in the present problem. The data was gathered from a single XGC simulation of shot 79688 of the JET tokamak. The range of the effective collision frequency of the points used is from $10^{-2} < \nu^* < 10$, where $\nu^*$ is defined as $\nu^* = \frac{\nu_ii}{qR/v_{thi}}$. The plasma parameters range from electron density $~9e19 m^3$ in the core to $~1e19 m^-3$ in the near scrape-off layer, and temperature $~5 keV$ in the core, to $~100 eV$ in the near scrape-off layer (for this discharge $T_e \approx T_i$). It may eventually be important to ensure the data spans a broad enough range of relevant physical parameters. Using the distribution function for both the ions and the electrons allows for calculation of the actual $\Delta f$ within the XGC collision kernel. The three sets of data, $f_i$, $f_e$, and $\Delta f_i$ can then be used for training the neural network according to the schematic depicted in Figure~\ref{fig:schematic}. Work is now underway to include $\Delta f_e$ in the training itself, rather than simply using the output $\Delta f_e$ from XGC's collision kernel.

Before training begins, a few steps are taken to preprocess the data. In the interest of reducing the training time, the training is limited to 4 toroidal planes. The data from these planes is filtered further to remove points that have nonphysical $f$ data and that have target $\Delta f$ that did not converge when passed into the XGC collision kernel. Here, nonphysical $f$ implies removing points that have negative values and unconverged points are points that have surpassed the imposed allowed number of iterations in the collision kernel. Once this filtering is done, the entire dataset is divided into three groups: training, validation, and testing, comprising 80\%, 10\%, and 10\% of the data respectively. The training data is then split into equal-sized batches before being passed into the neural network. 

Initial training was begun on U-Net \citep{Ronneberger2015}, one of the networks that performed best on semantic segmentation techniques earlier on. Once initial results were satisfactory, more sophisticated networks were investigated. In the end, the ReSeg network was chosen, with only a modification to the FC linear layer to account for the nature of the problem as regression, not classification (explained in Section 3). ReSeg combines Convolutional (CNN) and Recurrent Neural Networks (RNN) to perform semantic segmentation \citep{Visin2016}. ReSeg begins with a CNN, in this case VGG16 \citep{Simonyan2015}. This CNN downsamples the input distribution function, producing a grid that is 256 planes deep. ReSeg then stacks two layers of ReNet, each of which is composed of four RNNs that sweep over the distribution function. Two layers, interspersed with ReLU activations upsample the data to a 32 x 31 x 100 grid. For the purposes of this problem, the last layer of the network includes a convolutional kernel of size 1 x 1 x 100 with stride 1, which outputs the desired 32 x 31 x 1 distribution function representing $\Delta f$. For this problem, there are over 2.6 million parameters in the network.

\begin{figure}
    \centering
    \includegraphics[scale=0.38]{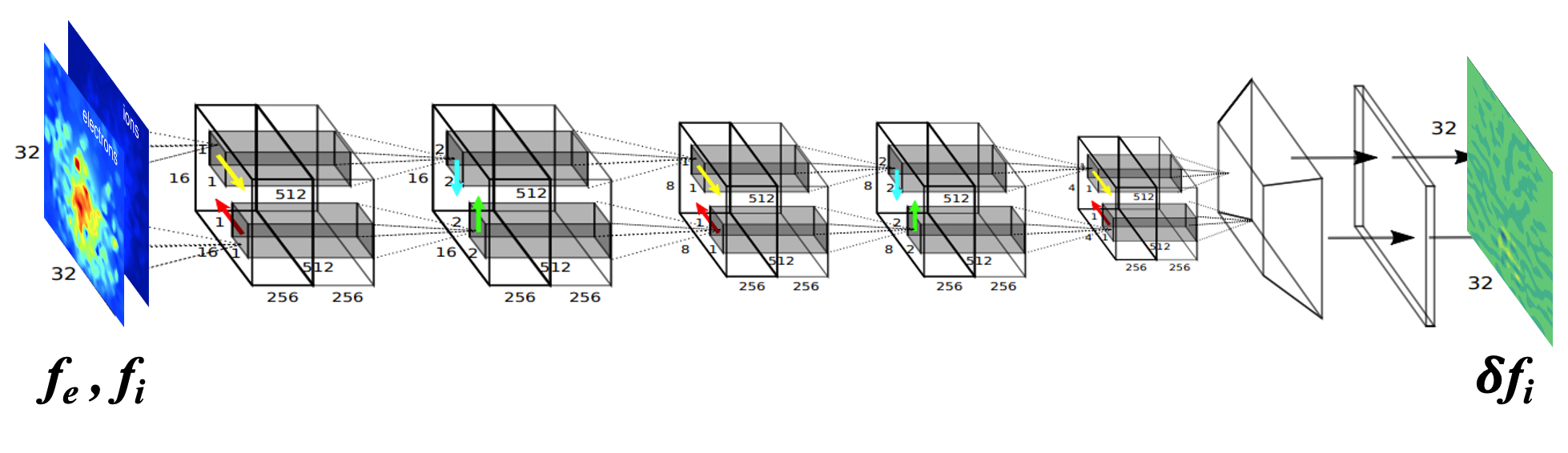}
    \caption{Schematic of ReSeg, a neural network architecture used for semantic segmentation, showing the an example input $f_e, f_i$ and output $\delta f_i$. The input $f_e$ and $f_i$ are concatenated to form a tensor of depth 2.}
    \label{fig:reseg}
\end{figure}

A schematic of the network architecture can be seen in Figure~\ref{fig:reseg}. Stochastic Gradient Descent (SGD) with momentum was used primarily as the optimizer, though RAdam, a variant of Adam that reduces the variance of adaptive learning rates \citep{Liu2019}, also showed success. Training was done for 100 epochs, with a validation set done twice an epoch. The training and validation results are shown in Figure~\ref{fig:training_overall}, with the $\ell_2$ and conservation error shown separately.

\section{Initial Training and Validation}

\begin{figure}
\begin{center}
    \begin{subfigure}{0.9\textwidth}
    \includegraphics[width=\textwidth]{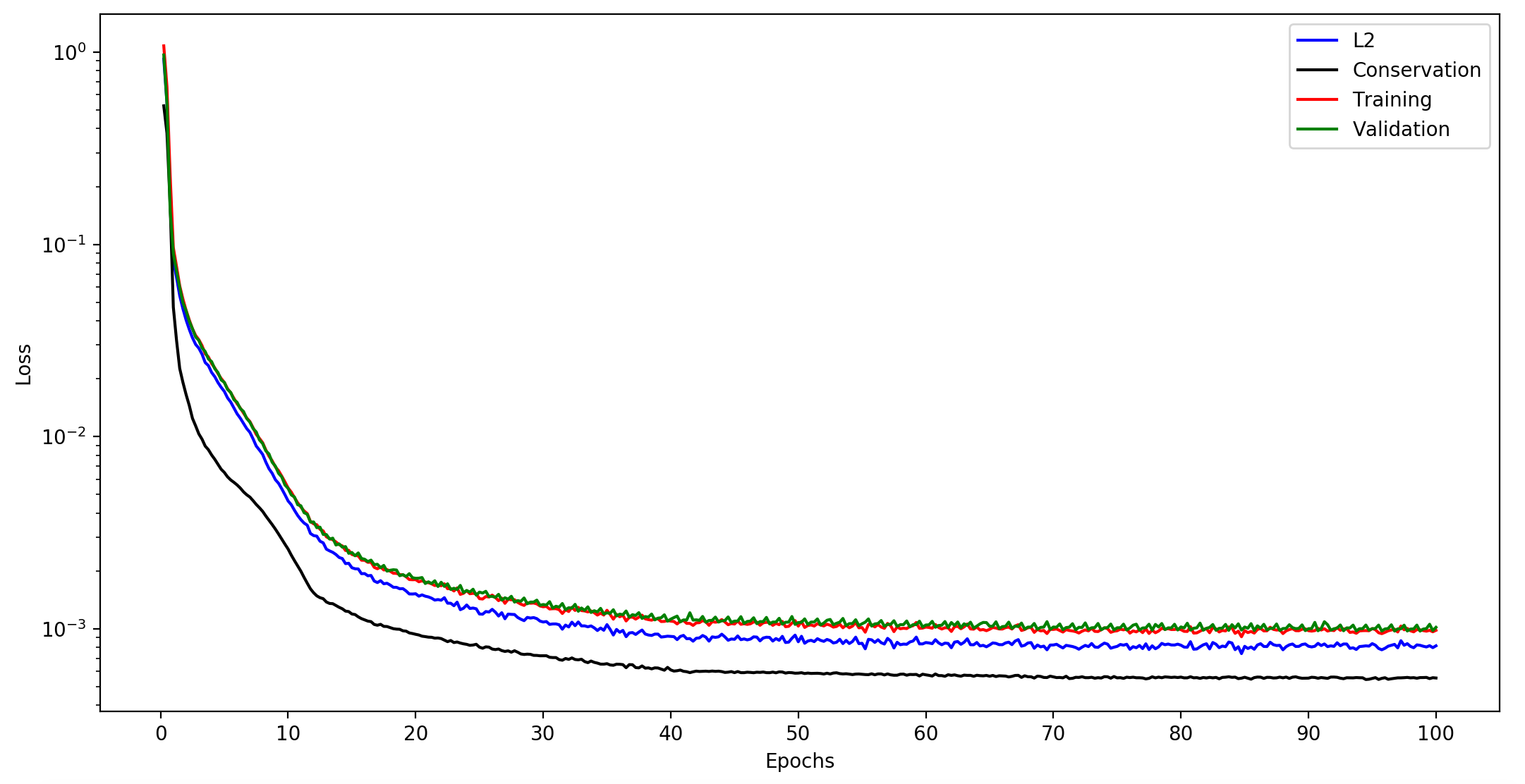}
    \label{fig:training_a}
\end{subfigure}

    \begin{subfigure}{0.9\textwidth}
    \includegraphics[width=\textwidth]{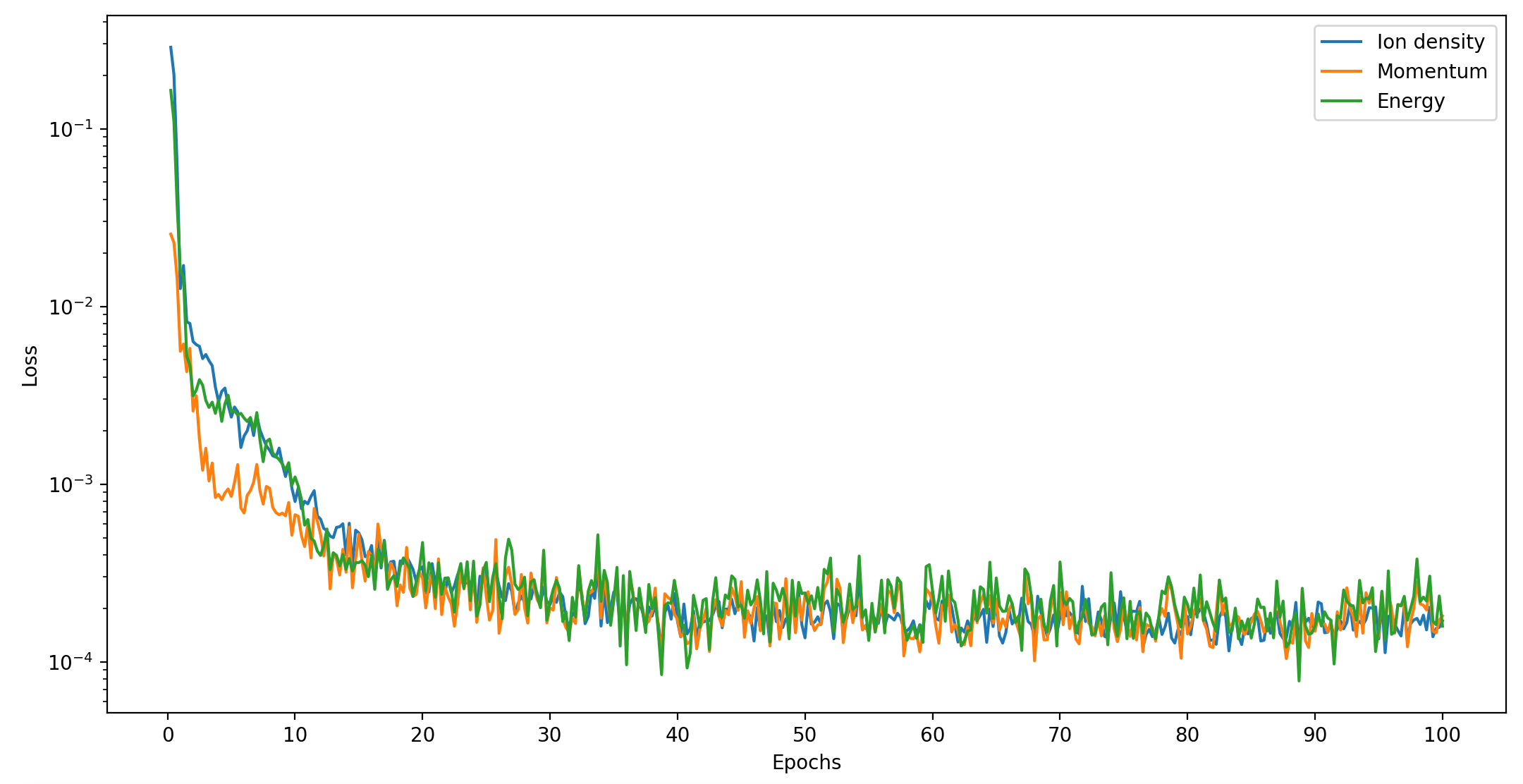}
    \label{fig:training_b}
    \end{subfigure}
  \caption{Training results on JET data for 100 epochs and four toroidal planes showing (a)  overall training and validation, as well as $\ell_2$ and conservation loss, and (b) the three components of the conservation loss separately.}
  \label{fig:training_overall}
\end{center}
\end{figure}

\begin{figure}
\begin{center}
  \includegraphics[width=0.8\textwidth]{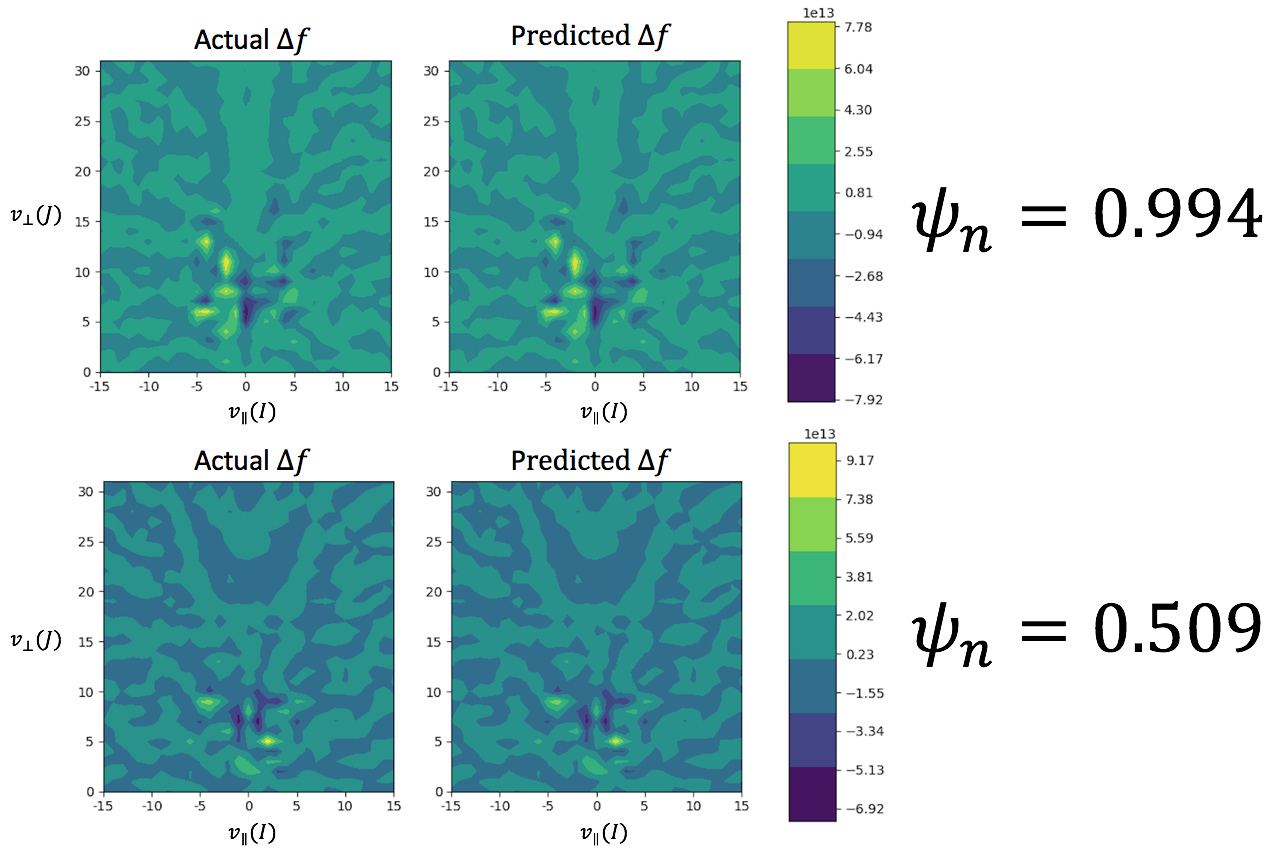}
  \caption{Comparison of $\Delta f_{col}$ prediction in two different regimes of collisionality. The x and y-axes represent bins in velocity space (parallel and perpendicular respectively). The index, $I$, representing $v_\parallel$ ranges from [-15, 15], totaling 31 bins in the parallel direction, and the index $J$, representing $v_\perp$ ranges from [0, 31], totaling 32 bins in the perpendicular direction.}
  \label{fig:collisionality}
\end{center}
\end{figure}

Figure~\ref{fig:training_overall} shows the training loss over time for 100 epochs of ReSeg training. The learning rate was adjusted using a combination of warmup and manual learning rate decay every 10 epochs. Note that the ``Training'' and ``Validation'' plots include the $\lambda_j$ from Equation \ref{eq:cons_loss}, while the ``Conservation'' plot does not. It is clear that after around 70 epochs, the loss begins to plateau and learning stagnates. This implores a finer tuning of the hyperparameters, or perhaps a revision of the optimization procedure, specifically in the weights assigned to each quantity in the loss function ($\lambda_j$ in Equation \ref{eq:cons_loss}). Figure~\ref{fig:training_overall}b breaks up the conservation loss into each of three conservation properties: ion density, momentum, and energy, the three elements of the summation in Equation \ref{eq:cons_loss}. Finally, Figure~\ref{fig:collisionality} illustrates the difference in learning of the structure of the $\Delta f$ for two different $\psi_n$, a normalized radial coordinate, which lie in two different regimes of collisionality. The lower the $\psi_n$, meaning the closer the plasma is to the core, the more Maxwellian the plasma, and so the smaller resulting $\Delta f$ from the collision operator. Collisions here result in a smaller $\Delta f_{col}$, perhaps making the transformation easier for the network to learn.

\begin{figure*}
\begin{center}
    \includegraphics[width=1\textwidth]{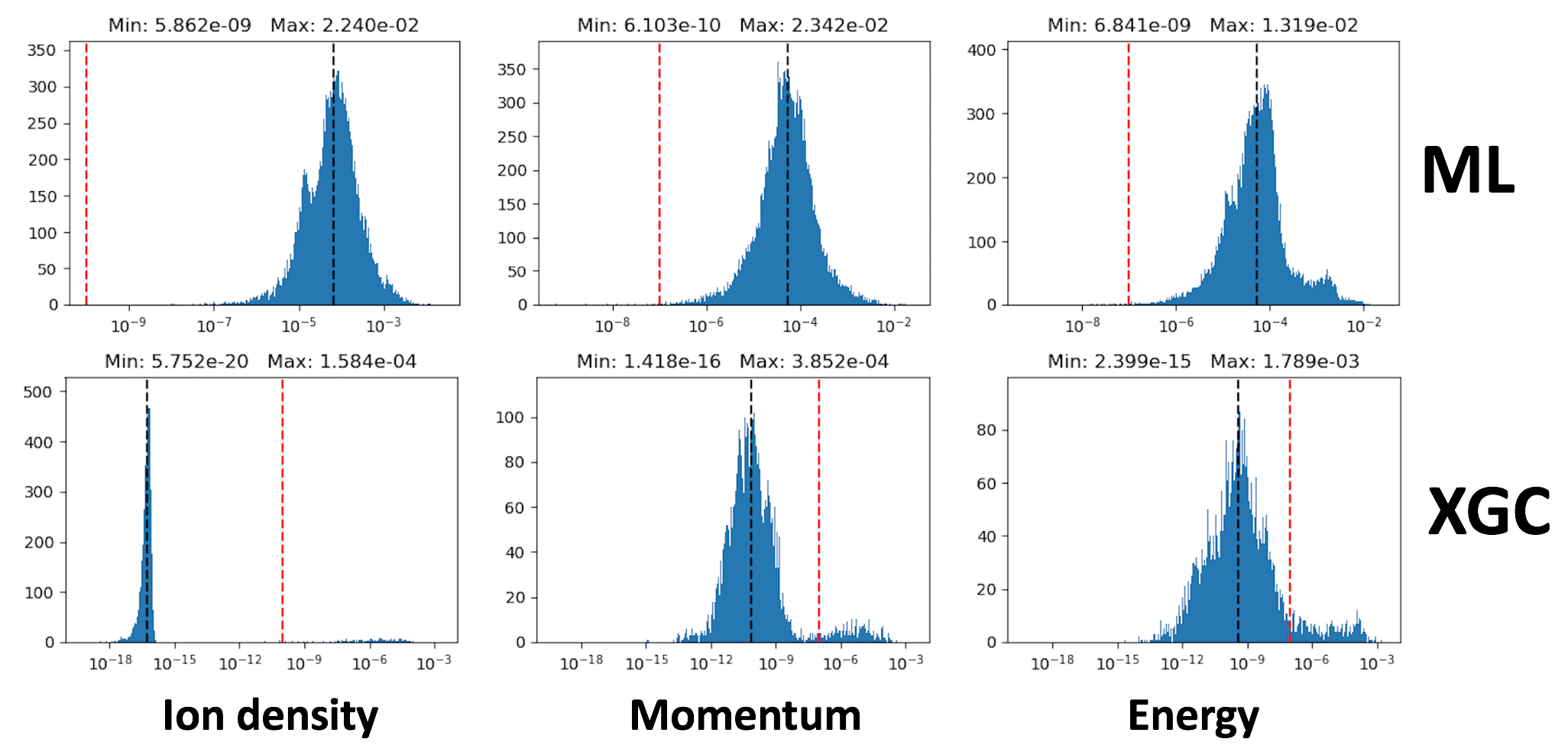}
    \captionof{figure}{Comparison of conservation properties from XGC and from test set of neural network using machine learning (ML). The black vertical dashed lines represent the median value for the specific quantity. The red ones represent the target value (see Equation \ref{eq:conv_crit}). Note: the x-axes (and y-axes) are different in both sets of plots.}
    \label{fig:fractions}
\end{center}
\end{figure*}

In order to declare success, it is necessary to look at how well the neural network can solve Equation \ref{eq:cons_eq}, i.e. how well it conserves mass, momentum, and energy. Looking at Equation \ref{eq:cons_loss}, we require each of the terms in the sum to be a minimum. Work still must be done to reduce the quantities in Figure~\ref{fig:fractions} to the threshold levels in XGC in case there is a drift type ML error propagation in time, but even these results show promise for reaching the precision required for the XGC code. The benefit of course to using the ML algorithm is that the time to solution should be much faster compared to the Picard iteration solver. Further work into optimization schemes has demonstrated even more promising results. In particular, an augmented Lagrangian scheme, which will be detailed in a subsequent report, has extended this initial ML application down to $10^{-6}$ accuracy.

\section{Conclusions and Further Work}

This work presents an encoder-decoder neural network that shows great potential in replacing the nonlinear FPL collision operator. Through a penalization method, the network enforces the relative conservation properties to $10^{-4}$ level, while minimizing $\ell_2$ error to $10^{-3}$ level (mimization of the conservation error is more important than the $\ell_2$ error minimization due to the random nature of the collision operation). Enforcement of the conservation constraints in the training have shown to greatly improve performance, and progress towards the goal of the low numerical error needed in a gyrokinetic code like XGC as demonstrated. 

Using the current trained model would result in a significant speedup of the collision kernel in the XGC code, especially with many plasma species. However, some improvement is needed to bring down the relative conservation accuracy to the $10^{-5}$ level for a safe application, with the time-integrated error limited to several percent, over thousands of XGC ion timesteps in case the error propagation is of drift type. The network has also proven successful in predicting this transformation for a wide range of training data that spans different regimes of collisionality. In short, this work has proven that machine learning can in fact serve as a useful aid for especially intricate calculations in numerical work. 

One immediate observation is that it would be useful to explore a wider variety of neural network architectures. It could be the case that an architecture that has perhaps not been as effective for image recognition problems could perform very well for the prediction of colliding plasma species. Furthermore, the optimization routine used here to impose the ``soft" constraints is fairly rudimentary and can certainly be refined to produce an even better trained network. Work to increase the sophistication of the optimization scheme in the form of an augmented Lagrangian method is underway and has shown promise in increasing the accuracy by two orders of magnitude. This work will be reported as a subsequent publication. Initial implementation of a trained ML collision operator model using this improved optimization method shows that it is five times faster than the currently used, highly optimized GPU version of the numerical solution of the integro-differential FPL collision operator. There still remain a number of ways to optimize the neural network that would improve this speedup even more.

On the physics side, it will be important in the future to investigate an even broader collisionality regime that might be anticipated in a detached divertor plasma. Regardless, training has already been done on plasma both in the core and in the edge, which already spans a wide range of collisionality. This can be seen in Figure~\ref{fig:collisionality}, which compares the prediction for points at two different places in the tokamak. 
It is also important to investigate the performance of the neural network for many species. 
Especially as the network begins to train on data from other plasma species, would new and separate training need to be done for different species? In the interest of reducing run-time, it is still necessary to make comparisons of run-time of using this neural network for inference in actual XGC production runs. All of these questions need to be addressed before the network can be expected to predict the $\Delta f$ reliably enough for use in the multispecies XGC collision kernel. 

\section*{Acknowledgements}
This work was supported in part by the U.S. Department of Energy, Office of Science, Office of Workforce Development for Teachers and Scientists (WDTS) under the Science Undergraduate Laboratory Internship (SULI) program. Support for this work was provided through the Scientific Discovery through Advanced Computing (SciDAC) program funded by the U.S. Department of Energy Office of Advanced Scientific Computing Research and the Office of Fusion Energy Sciences via the SciDAC-4 Partnership Center for High-fidelity Boundary Plasma Simulation (HBPS). This work was also supported by the U.S. Department of Energy, Office of Science, Office of Advanced Scientific Computing Research, Scientific Discovery through Advanced Computing (SciDAC) program via the FASTMath Institute under Contract No. DE-AC02-06CH11357 at Argonne National Laboratory and via the Partnership Center for High-fidelity Boundary Plasma Simulation at Princeton Plasma Physics Laboratory under the Contract No.DE-AC02–09CH11466. This research used resources of the Oak Ridge Leadership Computing Facility, which is a DOE Office of Science User Facility supported under Contract DE-AC05-00OR22725.

\bibliographystyle{jpp}

\bibliography{citations}

\end{document}